\documentclass[aps,floatfix,showpacs,nofootinbib,twocolumn]{revtex4-1}
\usepackage{amsmath}
\usepackage{amsfonts}
\usepackage{amssymb}
\usepackage{graphics,epsfig}
\usepackage{graphicx}
\begin{document}
\title{Visualizing Spacetime Curvature via Gradient Flows I: Introduction}
\author{Kayll Lake}
 \email{lake@astro.queensu.ca}
\affiliation{Department of Physics, Queen's University, Kingston,
Ontario, Canada, K7L 3N6 }

\date{\today}

\begin{abstract}
Traditional approaches to the study of the dynamics of spacetime curvature in a very real sense hide the intricacies of the nonlinear regime. Whether it be huge formulae, or mountains of numerical data, standard methods of presentation make little use of our remarkable skill, as humans, at pattern recognition. Here we introduce a new approach to the visualization of spacetime curvature. We examine the flows associated with the gradient fields of scalar invariants derived from the spacetime. These flows reveal a remarkably rich structure, and offer fresh insights, even for well known analytical solutions to Einstein's equations. The intent, however, is to go beyond idealized analytical solutions and eventually consider physically realistic situations. This requires a careful analysis of exactly which invariants that can actually be used in this approach. The present analysis serves as an overview and as an introduction to this program.
\end{abstract}
\maketitle

\section{Introduction}
To quote from a recent monograph \cite{gp}: \textit{``It appears that it is much easier to find a new solution of Einstein's equations than it is to understand it".} The present work involves the novel use of both computer algebra (for background calculations fundamental to the approach) and numerical routines (for numerical integration and visualization) with the objective being the development of a fresh view of ``curvature", and an consequent ``understanding" (for example, in the case of Einstein's theory) of a given spacetime. In this endeavor we are not alone. In \cite{dan}, a visualization of spacetime curvature, restricted to a study of the projected electric and magnetic components of the Weyl tensor, is being developed. In \cite{rez} a complementary approach is also under development. Here, we examine something akin to a highly complex dynamical systems approach: we look at the ``flows" associated with the gradient fields of scalar invariants of the background space. Preliminary results show that these flows reveal a remarkably rich structure and fresh insights even for well known geometries. The subject of invariants, even at dimension 4, is not closed in a mathematical sense. The approach considered here then necessarily imposes restrictions on the invariants used. We demand that any ``useable" invariant does not single out any particular observer (which requires the definition of ``observer independent" invariants, as explained below). The objects fundamental to \cite{dan} are also used here but in quite a different way: we use no projection. The approach we use does not, in a fundamental sense, use any particular theory. However, we also demand that the invariants used be connected directly to the underlying physics (and not offer just some general geometrical interpretation) and so we must adopt a particular theory to form this connection. We formulate the presentation given here within Einstein's theory of gravity.
\section{Gradient Vector Flows}
\subsection{Invariants}
Let $(M,g)$ be a semi - Riemannian manifold of dimension $d$ where $g$ is the metric tensor. The simplest construction of invariants involves scalars formed from metric contractions and partial derivatives of the metric tensor to order $p$ (to use the vernacular, ``scalars polynomial in the Riemann tensor").  The number ($\mathcal{N}$) of algebraically (not functionally) independent scalars of this type (invariants not satisfying polynomial degeneracies (syzygies), explained below) is given by \cite{thomas}
\begin{equation}\label{thomas}
    \mathcal{N}(d,p)=\frac{d(d+1)((d+p)!)}{2d!p!}-\frac{(d+p+1)!}{(d-1)!(p+1)!}+d
\end{equation}
for $d>2, p \geq 2$ and so, for example, in spacetime the number of non-differential invariants (no derivatives of the Riemann tensor) is $\mathcal{N}(4,2)=14$. Unfortunately, this does not mean that a single set of $14$ invariants will cover all possibilities. The situation is rather more involved \cite{stephani} and the mathematical issue of a minimal complete set remains incompletely resolved even at $d=4$. The problem is, perhaps, a problem of ``classical difficulty"\footnote{That is, unsolvable within the confines of the mathematical construction used.}.

Whereas the mathematics of polynomial invariants remains a very active area of research (see, for example, \cite{hervik} and references therein), as recently pointed out by Page \cite{page}, non-polynomial invariants can be constructed even when all polynomial invariants vanish. However, the invariants used by Page, in addition to extreme computational complexity, offer no clear physical interpretation and are not considered here.

When looking for exact solutions to Einstein's equations, many simplifications are made, and there is a consequent reduction in $\mathcal{N}$.  For example, in class $B$ warped product spacetimes (which includes, for example,  all spherical, plane and hyperbolic spacetimes) it is known that $\mathcal{N}(4,2)\leq4$ and that $\mathcal{N}(4,2)=1$ in the Ricci-flat case \cite{classb}. However, when one attempts more realistic situations, by way of numerical relativity, no symmetries or simplifying assumptions can be made and the value of $\mathcal{N}$ becomes unclear. Since the approach we are considering here should ultimately be useful in these more realistic spacetimes, we need to delineate the invariants that can be used. In this section we explain why there are exactly 8 polynomial invariants that can be used (for $d = 4$) to construct physically clear ``observer - independent" gradient flows. Going beyond these invariants explicitly introduces the four - velocity of the observer and so these invariants cannot be ``observer - independent". Our notation follows that in \cite{rg} and as explained there, the modern organization of polynomial invariants involves the construction of Ricci invariants (of which there are 4 independent invariants), Weyl invariants (again of which there are 4 independent invariants) and mixed invariants which involve contractions of the Ricci, Weyl (and dual) tensors, the number of which, for completion (that is, syzygy independence), is actually not known at present.

Let us start with the Ricci invariants. Einstein's equations are
\begin{equation}\label{einstein}
   R^{\alpha}_{\beta}-\frac{1}{2}\delta^{\alpha}_{\beta}R+\Lambda \delta^{\alpha}_{\beta} = 8 \pi T^{\alpha}_{\beta}
\end{equation}
where $R^{\alpha}_{\beta}$ is the Ricci tensor, $R$ the Ricci scalar, $\Lambda$ the cosmological constant and $T^{\alpha}_{\beta}$ the energy - momentum tensor. We use these equations to interpret the Ricci invariants. The first Ricci invariant is taken to be the Ricci scalar and we obtain
\begin{equation}\label{ricci}
    R=4\Lambda-8 \pi T
\end{equation}
where $T$ is the trace of $T^{\alpha}_{\beta}$. It is convenient, at this point, to use the trace - free Ricci tensor
\begin{equation}\label{tracefree}
    S^{\alpha}_{\beta} \equiv R^{\alpha}_{\beta}-\frac{1}{4}R\delta^{\alpha}_{\beta}
\end{equation}
so that Einstein's equations now take the form
\begin{equation}\label{traceeinstein}
    S^{\alpha}_{\beta} = 8 \pi (T^{\alpha}_{\beta}-\frac{T}{4}\delta^{\alpha}_{\beta}).
\end{equation}
The remaining three Ricci invariants are given by\footnote{The physically irrelevant numerical coefficients arise due to the translation from spinor notation to tensor notation, the notation we use in this introduction.}
\begin{equation}\label{r1}
    r_1 \equiv \frac{1}{4}S^{\alpha}_{\beta}S^{\beta}_{\alpha},
\end{equation}
\begin{equation}\label{r2}
    r_2 \equiv -\frac{1}{8}S^{\alpha}_{\beta}S^{\gamma}_{\alpha}S_{\gamma}^{\beta},
\end{equation}
and
\begin{equation}\label{r3}
    r_3 \equiv \frac{1}{16}S^{\alpha}_{\beta}S^{\gamma}_{\alpha}S^{\delta}_{\gamma}S^{\beta}_{\delta}.
\end{equation}
The physical interpretation of these invariants comes by way of (\ref{traceeinstein}). For example, it follows that
\begin{equation}\label{r1t}
    r_1=\frac{1}{4}(8 \pi)^2(T^{\alpha}_{\beta}T^{\beta}_{\alpha}-\frac{T^2}{4}),
\end{equation}
with analogous higher order expressions following for $r_2$ and $r_3$. With imposed symmetries, the number of algebraically independent Ricci invariants decreases.   For example, again in all class $B$ warped product spacetimes, and perhaps more, one can construct the syzygy \cite{classb}
\begin{equation}\label{syzygy}
    (-12r_3+7r_1^2)^3=(12r_2^2-36r_1r_3+17r_1^3)^2.
\end{equation}

The important points are the following: Ricci invariants $r_n$ with $n>3$ are not independent\footnote{In the general case one can algorithmically construct syzygies for all $r_n$ for $n>3$. The explicit forms for these syzygies up to $r_{10}$ are given in \cite{classb}.}, and the Ricci invariants are ``observer - independent". To explain this last point, note that for an observer with tangent 4 - vector $u^{\alpha}$, $u^{\alpha}$ does not enter the Ricci invariants. Whereas $T^{\alpha}_{\beta}$ can, in some cases (think of perfect fluids, for example), pick out a ``preferred" $u^{\alpha}$ (the phenomenological fluid streamlines), more general timelike 4 - velocities (``observers") simply do not enter the Ricci invariants. Indeed, for a perfect fluid, for example, the Ricci invariants reduce to polynomials (up to quartics) in the energy density and isotropic pressure.\footnote{This is an example of what we mean by giving a ``physical meaning" to an invariant.}  This invariance is, perhaps, made more poignant when we consider mixed invariants, as explained below. Next, however, we consider Weyl invariants for which ``observer - independence" is in fact manifest.

There are exactly 4 independent Weyl invariants (e.g. \cite{rg}) and these can be defined by\footnote{The notation used here is that $w1R$ and $w1I$ stand for the real and imaginary parts of the complex Weyl invariant $w1$, and similarly for $w2$. This notation is convenient. For example, the Kerr metric has only two independent invariants because it is of Petrov type D for which we always have the syzygy $6w2^2=w1^3$.}

\begin{equation}\label{w1r}
    w1R \equiv \frac{1}{8}C_{\alpha \beta \gamma \delta}C^{\alpha \beta \gamma \delta},
\end{equation}
where $C_{\alpha \beta \gamma \delta}$ is the Weyl tensor,
\begin{equation}\label{w1I}
    w1I \equiv \frac{1}{8}C^{*}_{\alpha \beta \gamma \delta}C^{\alpha \beta \gamma \delta},
\end{equation}
where $C^{*}_{\alpha \beta \gamma \delta}$ is dual to $C_{\alpha \beta \gamma \delta}$,
\begin{equation}\label{w2r}
    w2R \equiv -\frac{1}{16}{C_{\alpha \beta}}^{\gamma \delta}{C_{\gamma \delta}}^{\epsilon \zeta}{C_{\epsilon \zeta}}^{\alpha \beta}
\end{equation}
and
\begin{equation}\label{w2I}
    w2I \equiv -\frac{1}{16}{C^{*}_{\alpha \beta}}^{\gamma \delta}{C_{\gamma \delta}}^{\epsilon \zeta}{C_{\epsilon \zeta}}^{\alpha \beta}.
\end{equation}

Now define, in the usual way, the symmetric and trace - free ``electric" and ``magnetic" parts of the Weyl tensor:
\begin{equation}\label{electric}
    E_{\alpha \gamma} \equiv C_{\alpha \beta \gamma \delta}u^{\beta}u^{\delta}
\end{equation}
and
\begin{equation}\label{magnetic}
    B_{\alpha \gamma} \equiv C^{*}_{\alpha \beta \gamma \delta}u^{\beta}u^{\delta},
\end{equation}
where $u^{\alpha}$ is an \textit{arbitrary} unit timelike 4 - vector. Note that due to the symmetry properties of the Weyl tensor $E_{\alpha \beta}u^{\beta}=B_{\alpha \beta}u^{\beta}=0$. It can now be shown that in general (e.g. \cite{bell})
\begin{equation}\label{w1re}
    w1R = \frac{1}{16}(E_{\alpha \beta}E^{\alpha \beta}-B_{\alpha \beta}B^{\alpha \beta}),
\end{equation}
\begin{equation}\label{w1Ie}
    w1I = \frac{1}{8}(E_{\alpha \beta}B^{\alpha \beta}),
\end{equation}
\begin{equation}\label{w2re}
    w2R = \frac{1}{32}(3E^{\alpha}_{\beta}B^{\gamma}_{\alpha}B_{\gamma}^{\beta}-E^{\alpha}_{\beta}E^{\gamma}_{\alpha}E_{\gamma}^{\beta})
\end{equation}
and
\begin{equation}\label{w2Ie}
    w2I = \frac{1}{32}(B^{\alpha}_{\beta}B^{\gamma}_{\alpha}B_{\gamma}^{\beta}-3E^{\alpha}_{\beta}E^{\gamma}_{\alpha}B_{\gamma}^{\beta}).
\end{equation}

The ``observer - independent" nature of the 4 Weyl invariants follows from the fact that these invariants are independent of the  unit timelike 4 - vector (``observer") chosen to split the Weyl tensor into its electric and magnetic parts. To properly understand exactly what this means, we must move on to mixed invariants.\footnote{At this point, one might ask what happened to the familiar Kretschmann scalar $R_{\alpha \beta \gamma \delta}R^{\alpha \beta \gamma \delta}$ where $R_{\alpha \beta \gamma \delta}$ is the Riemann tensor. The answer is that Kretschmann is no longer considered fundamental. It is given by $R^2/6+8(r1+w1R)$. } First, however, it is appropriate to compare the objects used in \cite{dan}. There, projections are used so that the fundamental objects of interest are
\begin{equation}\label{et}
    \mathcal{E}_{\alpha \beta} = h_{\alpha}^{\gamma}h_{\beta}^{\delta}E_{\gamma \delta}
\end{equation}
and
\begin{equation}\label{bt}
    \mathcal{B}_{\alpha \beta} = -h_{\alpha}^{\gamma}h_{\beta}^{\delta}B_{\gamma \delta}
\end{equation}
where $h$ is the projection tensor
\begin{equation}\label{h}
    h^{\alpha}_{\beta}\equiv\delta^{\alpha}_{\beta} + u^{\alpha}u_{\beta}.
\end{equation}
The physical meaning of both $\mathcal{E}_{\alpha \beta}$ and $\mathcal{B}_{\alpha \beta}$ is explored in considerable detail in \cite{dan}.\footnote{To be precise, the $3$ by $3$ quantities $\mathcal{E}_{i j}$ and $\mathcal{B}_{i j}$ are used. These are obtained after projection onto the spatial slices using $h^{\alpha}_{\beta}$.} In very quick summary, $\mathcal{E}$ produces tidal forces and $\mathcal{B}$ produces differential frame dragging. It is precisely because of these physical interpretations that one introduces the splitting of the Weyl tensor in the first place. But, how then do $\mathcal{E}_{\alpha \beta}$ and $E_{\alpha \beta}$ and $\mathcal{B}_{\alpha \beta}$ and $B_{\alpha \beta}$ differ?  A straightforward calculation gives
\begin{equation}\label{ete}
    \mathcal{E}_{\alpha \beta}= E_{\alpha \beta}, \;\; \mathcal{B}_{\alpha \beta}= -B_{\alpha \beta},
\end{equation}
and so we are considering the invariants built out of the same fundamental objects used in \cite{dan}, but we are using these objects in quite a different way.

The next invariant in, the scheme of \cite{rg}, is the complex mixed invariant $m1$. The real and imaginary parts of this invariant are given by
\begin{equation}\label{m1r}
    m1R \equiv \frac{1}{8}C_{\alpha \beta \gamma \delta}S^{\alpha \gamma}S^{\beta \delta}
\end{equation}
and
\begin{equation}\label{m1I}
    m1I \equiv \frac{1}{8}C^{*}_{\alpha \beta \gamma \delta}S^{\alpha \gamma}S^{\beta \delta}
\end{equation}
respectively. Splitting the Weyl tensor as before we now obtain \cite{classb}
\begin{equation}\label{m1rs}
    m1R =\frac{1}{8}(2E^{\alpha }_{\beta} S^{\beta}_{\delta} S^{\delta}_{\alpha} - 2 E_{\alpha \beta} S^{\alpha \beta} S_{\gamma \delta}
  u^{\gamma} u^{\delta} + S^{\alpha}_{\gamma} u^{\gamma} E_{\alpha \beta} S^{\beta}_{\delta} u^{\delta})
\end{equation}
and
\begin{equation}\label{m2rs}
    m1I =\frac{1}{8}(2B^{\alpha }_{\beta} S^{\beta}_{\delta} S^{\delta}_{\alpha} - 2 B_{\alpha \beta} S^{\alpha \beta} S_{\gamma \delta}
  u^{\gamma} u^{\delta} + S^{\alpha}_{\gamma} u^{\gamma} B_{\alpha \beta} S^{\beta}_{\delta} u^{\delta}).
\end{equation}
Now, something new has entered the picture. We now have the vector
\begin{equation}\label{vector}
    S^{\alpha}_{\beta}u^{\beta}
\end{equation}
and scalar
\begin{equation}\label{scalar}
    S_{\alpha \beta}u^{\alpha}u^{\beta}
\end{equation}
to deal with.\footnote{At first, these terms may seem very strange. To be explicit here, they arise from the first term on the right hand side of equation (3.36) in the first reference in \cite{stephani}. } These bring our particular choice of $u^{\alpha}$ into the picture and so the mixed invariant $m1$ is not ``observer - independent". One option is to simply avoid the splitting of the Weyl tensor. This is not viable because it is precisely through this splitting that one eventually obtains the physical meaning of the Weyl tensor, as explained above. Another option is to allow $m1$ and higher invariants\footnote{All higher order invariants necessarily include terms which depend on $u^{\alpha}$.} into the study of gradient flows. We do not see this as a viable option either since the whole approach used here is to reveal observer independent properties of the spacetime. The only viable option seems to be, at least for polynomial invariants, a restriction to the 4 Ricci invariants and 4 Weyl invariants (at dimension $d=4$) as regards the construction of gradient flows.

\subsection{Vector Fields}
Given $(M,g)$ and a set of invariants $\mathcal{I}_{n}$ (we need not specify $d$ nor $p$ at this point nor even the nature of the $\mathcal{I}_{n}$) consider the gradient flows\footnote{The choice $``-"$,  a negative gradient flow, is in keeping with modern mathematical convention. The use of $``\pm"$ would require a more involved classification of critical points in odd dimensions (explained below) which we avoid. Here $\nabla$ is the covariant derivative and $\partial $ the partial derivative.} \cite{katok}
\begin{equation}\label{gradient}
    k^{\alpha}_{n} \equiv - \nabla ^{\alpha}\mathcal{I}_{n} = - g^{\alpha \beta}\frac{\partial \mathcal{I}_{n}}{\partial x^{\beta}}.
\end{equation}
 Note that $k$ can be timelike, spacelike or null. Also note that even if a flow resides entirely in a subspace, calculations must be carried out in the full space since invariants of the full space are not, of course, the same as invariants of subspaces. For notational simplicity we now consider one invariant at a time and so drop $n$.
 Let us write the gradient flow for a particular invariant in the form
\begin{equation}\label{dynamic}
    k^{\alpha} = \dot{x}^{\alpha}=- g^{\alpha \beta}\frac{\partial \mathcal{I}}{\partial x^{\beta}}
\end{equation}
where $^{.} \equiv d/d\lambda$ and $\lambda$ is any natural parametrization of the curve $x^{\alpha}(\lambda)$ with tangent $k^{\alpha}$. It is clear that we are dealing with an autonomous dynamical system. First, however, let us gather together some fundamental properties of $k$.
 By construction, we have
\begin{equation}\label{gradient1}
    \nabla_{[\alpha}k_{\beta]}= k_{[\alpha}\nabla_{\beta}k_{\gamma]}=0
\end{equation}
and so on. In terms of the norm
\begin{equation}\label{norm}
    k_{\alpha}k^{\alpha} \equiv \mathcal{V},
\end{equation}
it follows from (\ref{gradient1}) that the acceleration associated with $k$ is\footnote{It is essential that $k$, like a Killing field, not be normalized, since with (\ref{acceleration}), normalization would restrict $k$ to a geodesic flow.}
\begin{equation}\label{acceleration}
    k^{\beta}\nabla_{\beta}k_{\alpha}=\frac{1}{2}\nabla_{\alpha}\mathcal{V}.
\end{equation}
From (\ref{gradient}) it follows that the associated expansion is
\begin{equation}\label{expansion}
    \nabla_{\alpha} k^{\alpha}= - \nabla_{\alpha}(\nabla^{\alpha} \mathcal{I}) \equiv  - \square \;\mathcal{I}.
\end{equation}
Borrowing the usual definition of the vorticity tensor $\omega_{\alpha \beta} \equiv h_{\alpha}^{\gamma}h_{\beta}^{\delta}\nabla_{[\gamma}k_{\delta]}$, it follows immediately from (\ref{gradient1}) that the flows we are considering are irrotational,
\begin{equation}\label{rot}
    \omega_{\alpha \beta}=0.
\end{equation}

The gradient flows (\ref{gradient}) can encounter a variety of exceptional situations, the most common of which are critical ``points" were $k^{\alpha}=0$. This is discussed below. Perhaps the simplest exceptional situation is the development of a caustic where the expansion of the flow diverges so that, from (\ref{expansion}),
\begin{equation}\label{caustic}
    |\square \;\mathcal{I}| \rightarrow \infty.
\end{equation}

In a recent influential work \cite{vfw}, it has been claimed that attempts to explain cosmic acceleration, by way of observations in locally inhomogeneous spacetimes, necessarily involves a local ``weak singularity" at the origin.  This argument is based on the observation that $\square \; R$ necessarily diverges at the origin in these models.  This point of view has been criticized \cite{bkhc}.  In terms of a gradient flow, as introduced here, the correct way to view this controversy is to observe that this ``weak singularity" is actually a caustic in the gradient flow of $R$, as follows from (\ref{expansion}). Since the models under consideration are simply dust, $k$ represents the gradient flow of the energy density. The ``weak singularity" derives from the lack of sufficiently high differentiability of the energy density at the origin. More substantial, and certainly more well known singularities, occur where any $\mathcal{I}$ diverges (so called ``scalar polynomial singularities").
\subsection{Special Symmetries}
For any scalar field $\mathcal{I}$, the Lie derivative associated with a vector field $\xi$ is given by \cite{yano}
\begin{equation}\label{lie}
\mathcal{L}_{\xi} \mathcal{I}=\xi_{\alpha} \nabla ^{\alpha} \mathcal{I} = \xi_{\alpha} k^{\alpha}.
\end{equation}
In certain circumstances, the scalar $\xi_{\alpha} k^{\alpha}$ can be directly related to the invariant $\mathcal{I}$ itself. For example, if the manifold admits a homothetic motion, $\mathcal{L}_{\xi}g_{\alpha \beta}=2 c g_{\alpha \beta}$, where $c$ ($=\nabla_{\alpha}\xi^{\alpha}/d$) is constant, it is known that for polynomial invariants \cite{pelavas}
\begin{equation}\label{pelavas}
    \mathcal{L}_{\xi} \mathcal{I}=\kappa c \mathcal{I}
\end{equation}
where $\kappa$ is an integer characteristic of $\mathcal{I}$ (involving, for example, $p$ and the number of discrete contractions used to make $\mathcal{I}$). In the case that $\xi$ is Killing ($c=0$) we then have
\begin{equation}\label{killing}
    \xi_{\alpha}k^{\alpha}=0.
\end{equation}
That is, \textit{polynomial gradient flows are orthogonal to Killing flows}, should they exist.\footnote{In the special circumstance that $k$ itself satisfies $\mathcal{L}_{k}g_{\alpha \beta}=2 c g_{\alpha \beta}$, it follows that $k$ is concurrent \cite{yano}, $\nabla_{\alpha}k_{\beta}=\Phi g_{\alpha \beta}$, and so the associated streamlines are geodesic with $\nabla_{\alpha}\mathcal{V}=2\Phi k_{\alpha}.$} This result can be used to clarify and extend the classical notions of $\mathcal{R}$ and $\mathcal{T}$ regions of spacetime. This is explored in Appendix A.

\section{Dynamical Systems}

\subsection{Classification of Critical Points\footnote{In this section we do not designate the dimension $d\geq2$ simply because we do not need to.}}

The phase portrait associated with covariant counterpart to (\ref{dynamic}) is dominated by critical points ($\mathcal{P}$, where $k^{\alpha}=k_{\alpha}=0$\footnote{It goes without saying, but we say it here anyways, that the existence of a critical point is a coordinate independent feature of a flow, up to degenerate coordinates which do not distinguish isolated critical points. An example of degenerate coordinates is given below.}), the deep study of which leads to Morse Theory \cite{jost}. Gradient systems, in the large, have many diverse uses, and are the subject of much modern mathematical study. In this section, we gather together the information required to simply \textit{classify} critical points of gradient systems. It is to be noted that in many presentations, there is no mention of covariant nor contravariant components since there is in the usual context no difference. This is certainly not true here. When we talk about phase portraits these are associated with the covariant field $k_{\alpha}$. When we talk about critical points $\mathcal{P}$ we mean $k^{\alpha}=k_{\alpha}=0$.

Clearly the gradient
\begin{equation}\label{gradients}
   \frac{\partial \mathcal{I}}{\partial x^{\beta}} = - k_{\beta}
\end{equation}
is a covariant vector (and that is why we talk about phase portraits of $k_{\alpha}$) and so the standard Hessian
\begin{equation}\label{shessian}
    \frac{\partial^2 \mathcal{I}}{\partial x^{\beta} \partial x^{\alpha}} = - \frac{\partial k_{\beta}}{\partial x^{\alpha}}
\end{equation}
is not covariant. Rather, we use the covariant Hessian
\begin{equation}\label{chessian}
    H_{\alpha \beta} \equiv - \nabla_{\alpha} k_{\beta}
\end{equation}
which, from (\ref{gradient1}), is symmetric with the trace given by (\ref{expansion}). (Of course at critical points, the covariant Hessian reduces to the ordinary Hessian.) Let $H$ designate the determinate of $H_{\alpha \beta}$, most often called the ``discriminant". Under coordinate transformations $\bar{x}^{\alpha}(x^{\beta})$,
\begin{equation}\label{detH}
    \bar{H} = \mathcal{J}^2H,
\end{equation}
where $\mathcal{J}$ (we assume $\neq 0$) is the Jacobian of the transformation. As a result, the sign of $H$ is invariant to coordinate transformations (a condition central to what follows). A critical point $\mathcal{P}$  is degenerate if the discriminant vanishes, $H|_{\mathcal{P}}=0$. Otherwise $\mathcal{P}$ is non-degenerate, that is, a Morse critical point.

Next, we must characterize the local maxima and minima of $\mathcal{I}$ at $\mathcal{P}$, viewed as functions $\mathcal{I}(x^{\alpha})$. This is done by way of a covariant generalization of the usual procedure (see, for example, \cite{apostol}). Let $\Delta_{d-\alpha}$ be the cofactors $C(\alpha,\alpha)$ of $H_{\alpha \beta}$. Define $\Delta_{0}=1$ and create the list $\Delta_{0},...\;,\Delta_{d}$. If all $d+1$ cofactors are positive, then $\mathcal{I}$ has a local minimum at $\mathcal{P}$. If they alternate in sign, $\mathcal{I}$ has a local maximum at $\mathcal{P}$. Otherwise, the character of $\mathcal{I}$ at $\mathcal{P}$ is undetermined.

We are now in a position to summarize the situation \cite{hale}: For isolated Morse critical points:
\begin{itemize}
\item $\mathcal{P}$ is an unstable node if and only if $H|_{\mathcal{P}}>0$ and $\mathcal{I}$ has a local maximum at $\mathcal{P}$.
\item $\mathcal{P}$ is an asymptotically stable node if and only if  $H|_{\mathcal{P}}>0$ and and $\mathcal{I}$ has a local minimum at $\mathcal{P}$.
\item $\mathcal{P}$ is a saddle point if and only if  $H|_{\mathcal{P}}<0$.
\end{itemize}
It is to be noted that gradient flows are simple flows in the sense that critical points are the only possible limit sets and that for gradient flows all critical points are hyperbolic. Moreover, if the critical points are all isolated, the trajectories of the flow all terminate at critical points (for all of these points see, for example, \cite{katok}). Since flows can, and are in simple cases, restricted to subspaces, if $H|_{\mathcal{P}}=0$ for the full space, we classify the critical points for $H|_{\mathcal{P}}$ evaluated in the subspace associated with the flow. An example is given below.

\subsection{Index Theory}
Let us start with 2-dimensional flows and characterize the topology associated with the flow \cite{perko}.
We note the following regarding the Poincar\'{e} index:
\begin{itemize}
\item The index of a node is $+1$.
\item The index of a source or sink is $+1$.
\item The index of a hyperbolic saddle point is $-1$.
\item The index of a closed curve containing fixed points is equal to the sum of the indices of the fixed points within.
\end{itemize}

The importance of calculating theses indices follows from the Poincar\'{e} - Hopf theorem which we state here in the following simple (and unrigorous) form: Suppose the flow is on a 2-surface $\Sigma$. Calculate the sum of the indices of all critical points on $\Sigma$ with a suitable curve $\gamma$ which contains all isolated critical points. Then this sum is the Euler Characteristic of $\Sigma$ within $\gamma$. Higher dimensions follow analogously (see \cite{katok}). To summarize: the detailed characteristics of the isolated critical points of a gradient flow are governed by the topology of the background space supporting the flow.

\section{Newtonian Analogues}

In this section we examine invariants that allow the construction of Newtonian analogues. Consider here the simplest case, purely electric spacetimes $B_{\alpha \beta}=0$ (e.g. all static spacetimes, shear-free and hypersurface orthogonal perfect fluid spacetimes \textit{etc.}). The tidal tensor is defined by (e.g. \cite{hobson})
 \begin{equation}\label{tider}
    \mathcal{T}_{\alpha \gamma} \equiv R_{\alpha \beta \gamma \delta}u^{\beta}u^{\delta}
 \end{equation}
where $u^{\alpha}$ is a unit timelike 4 - vector. With the aide of (\ref{electric}) it follows that the tidal scalar $\mathcal{T}$ is given by
\begin{equation}\label{stider}
    \mathcal{T} \equiv \mathcal{T}_{\alpha \beta}\mathcal{T}^{\alpha \beta} = E_{\alpha \beta}E^{\alpha \beta} + \Omega
\end{equation}
where $\Omega$ contains terms involving $R_{\alpha \beta}u^{\alpha}u^{\beta}$. Since we do not allow these observer dependent terms, our considerations here are restricted to the Ricci - flat case. We now construct the gradient flow
\begin{equation}\label{tidalflow}
        k_{\gamma} \equiv -\nabla_{\gamma}(E_{\alpha \beta}E^{\alpha \beta}).
\end{equation}

In a Euclidean three - space with coordinates $x^{a}$ and metric tensor $\eta_{a b}$ construct the potential flow
\begin{equation}\label{newtonflow}
    l_{a}=-\frac{\partial \Phi}{\partial x^{a}}
\end{equation}
where $\Phi$ the Newtonian potential. The Newtonian tidal tensor is the trace - free covariant Hessian (e.g. \cite{ehlers})
\begin{equation}\label{tide}
    E_{a b}=-\nabla_{a}l_{b}-\frac{1}{3}\eta_{a b} \square \;\Phi
\end{equation}
from which we construct the associated invariant
\begin{equation}\label{newinvar}
    E_{a b}E^{a b}=\nabla_{a}l_{b}\nabla^{a}l^{b} -\frac{1}{3}(\square \;\Phi)^2.
\end{equation}
We now construct the gradient flow
\begin{equation}\label{fields}
   k_{c} \equiv - \nabla_{c}(E_{a b}E^{a b}).
\end{equation}
The covariant derivatives in (\ref{tide}), (\ref{newinvar}) and (\ref{fields}) are with respect to $\eta_{a b}$.

We say that the flow (\ref{fields}) is a Newtonian analogue (in no way any limit) of the flow (\ref{tidalflow}) (once projected onto a spatial 3 - slice) if their associated phase portraits are ``analogous". By ``analogous" we refer to, for example, the topology of the flow. The Curzon - Chazy solution provides a non-trivial detailed demonstration, and clarification, of these ideas \cite{curzon}.

\section{Elementary Examples}
\subsection{The Robertson-Walker Spacetime}
Perhaps the most widely recognized spacetime is that of Robertson and Walker,
\begin{equation}\label{rw}
    ds^2=a(t)^2(\frac{dr^2}{1-kr^2}+r^2 d \Omega_2^2)-dt^2.
\end{equation}
Due to the spatial isotropy, it is clear, \textit{a priori}, that the associated gradient ``flows" can only be 1-dimensional.  There can be no isolated critical points. The negative gradient flows are given by
\begin{equation}\label{gradrw}
    k^{\alpha} = \frac{d \mathcal{I}}{d t} \delta^{\alpha}_{t}
\end{equation}
so that
\begin{equation}\label{Krw}
    \mathcal{V}=-\left(\frac{d \mathcal{I}}{d t}\right)^2 \leq 0.
\end{equation}
From (\ref{ricci}) and (\ref{r1t}) it follows that for any perfect fluid
\begin{equation}\label{perfect}
    8 \pi \rho = \sqrt{3 r_1}+\frac{R}{4}-\Lambda,
\end{equation}
where $\rho$ is the energy density, and so taking
\begin{equation}\label{rwi}
    \mathcal{I} = \sqrt{3 r_1}+\frac{R}{4}-\Lambda = \frac{3}{a^2}(\dot{a}^2+k)-\Lambda,
\end{equation}
where $^{.} \equiv d/dt$, it follows that
\begin{equation}\label{rwk}
    k_{\alpha}=\frac{6}{a^3}\dot{a}(k+\dot{a}^2-a \ddot{a})\delta_{\alpha}^{t}.
\end{equation}
Turn - around Universes, those for which $\dot{a}(t_{0})=0$, define critical spacelike 3-surfaces $t=t_{0}$. These 3-surfaces are degenerate in the sense that at $t_{0}$
\begin{equation}\label{rwdegen}
    k_{\alpha} = \mathcal{V}=\nabla_{\alpha}k^{\alpha}=k^{\beta} \nabla _{\beta} k_{\alpha}=0.
\end{equation}

\subsection{The Schwarzschild Spacetime}
Consider next the Schwarzschild vacuum. This is a significant step upward in complexity over the Robertson - Walker case. In dimensionless null coordinates $(u,v,\theta,\phi)$ the solution appears as
\begin{equation}\label{kruskal}
ds^2=4m^2\left(\frac{4 \mathcal{L}}{(1+\mathcal{L})}\frac{dudv}{uv}+(1+\mathcal{L})^2 d \Omega^2_{2}\right),
\end{equation}
where $m>0$ is the effective gravitational mass, $\mathcal{L}\equiv \mathcal{L}(uv)$ and $\mathcal{L}$ is the Lambert W function \cite{lambert} \cite{kruskal}. (It remains convenient to retain the ``Schwarzschild $r$" defined by $r \equiv 2m(1+\mathcal{L})$.)
It follows that $\xi^{\alpha}\equiv(u,-v,0,0)$ is a Killing vector (and tangent to trajectories of constant``Schwarzschild $r$"). We find
\begin{equation}\label{xikruskal}
\xi^{\alpha}\xi_{\alpha}=\frac{(4m)^2}{r}(2m-r), \xi^{\alpha}\nabla_{\alpha}\xi^{\beta}=(\frac{2m}{r})^2(u,v,0,0)
\end{equation}
and so along $u=0$ or $v=0$ ($r=2m$), $\xi$ is tangent to a radial null geodesic. Further, we find that $\nabla_{\alpha}\xi^{\alpha}=\xi_{[\alpha}\nabla_{\beta}\xi_{\gamma]}=0$ so that regions for which $r>2m$ are static; all of which is, of course, well known.

For (\ref{kruskal}), there is only one independent invariant of order $p=2$, and we can take this to be the square of the Weyl tensor ($8 w1R$),
\begin{equation}\label{invarschw}
    \mathcal{I}=\frac{3}{4 m^4 (1+\mathcal{L})^6}.
\end{equation}
It follows that the tangents to the associated negative gradient flows are given by
\begin{equation}\label{kruskalflowup}
    k^{\alpha} =\chi (u,v,0,0)
\end{equation}
where
\begin{equation}\label{psi}
    \chi \equiv \left(\frac{3}{4 m^3}\right)^2\frac{1}{(1+\mathcal{L})^7}
\end{equation}
and
\begin{equation}\label{kruskalflow}
    k_{\alpha} =\psi (\frac{1}{u},\frac{1}{v},0,0)
\end{equation}
where
\begin{equation}\label{psi}
    \psi \equiv \frac{9\mathcal{L}}{2m^4(1+\mathcal{L})^8}.
\end{equation}
Both $\xi$ and $k$ vanish identically at the isolated critical point $\mathcal{P}$, the bifurcation 2-sphere $u=v=0$.\footnote{Whereas it is obvious that $k^{\alpha}$ vanishes at $u=v=0$, to see that $k_{\alpha}$ also vanishes there note that $\lim_{u \rightarrow 0} \mathcal{L}(uv)/u=v $ and $\lim_{v \rightarrow 0}\mathcal{L}(uv)/u=u $. } The use of a complete covering of the manifold \textit{ab initio} here is crucial as regards isolated critical points. If, for example, one was to use traditional ``Schwarzschild" coordinates $(r,\theta,\phi,t)$ then the span $(r=2m, -\infty <t <\infty)$ is critical in the sense that $k^{\alpha}=\xi^{\alpha}=0$ over this span which is, of course, but a subset of the bifurcation two-sphere. Next, we find the norm
\begin{equation}\label{vkruskal}
   \mathcal{V} = \left(\frac{9}{4}\right)^2\frac{\mathcal{L}}{m^{10}(1+\mathcal{L})^{15}} \propto \frac{r-2m}{r^{15}}.
\end{equation}
 The $\mathcal{R}$ region corresponds to $r>2m$ whereas the $\mathcal{T}$ region corresponds to $r<2m$, exactly as expected. For the expansion we find
 \begin{equation}\label{kruskalexpansion}
    \nabla_{\alpha}k^{\alpha}=-\left(\frac{9}{8}\right)\frac{5 \mathcal{L}-1}{m^6(1+\mathcal{L})^9} \propto \frac{5r-12m}{r^9}.
 \end{equation}
 The acceleration is given by
 \begin{equation}\label{acckruskal}
    k^{\alpha} \nabla_{\alpha}k^{\beta} = -\delta(u,v,0,0)
\end{equation}
 where
\begin{equation}\label{delta}
    \delta =  \left(\frac{9}{16}\right)^2\frac{14 \mathcal{L}-1}{m^{12}(1+\mathcal{L})^{16}} \propto \frac{7r-15m}{r^{16}}.
\end{equation}
From (\ref{kruskalflowup}) and (\ref{acckruskal}) it follows that along $u=0$ or $v=0$, $k$ is tangent to a radial null geodesic.

\begin{figure}[ht]
\epsfig{file=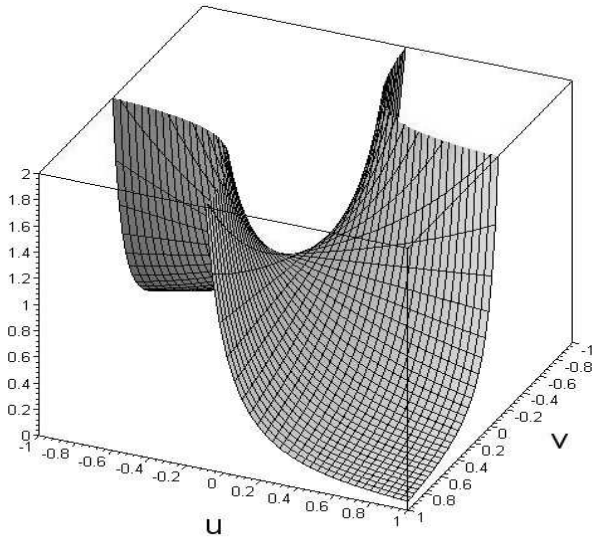,height=3in,width=3.5in,angle=0}
\caption{\label{schw1}The function  $m^4\mathcal{I}$ for the Schwarzschild spacetime.}
\end{figure}

\begin{figure}[ht]
\epsfig{file=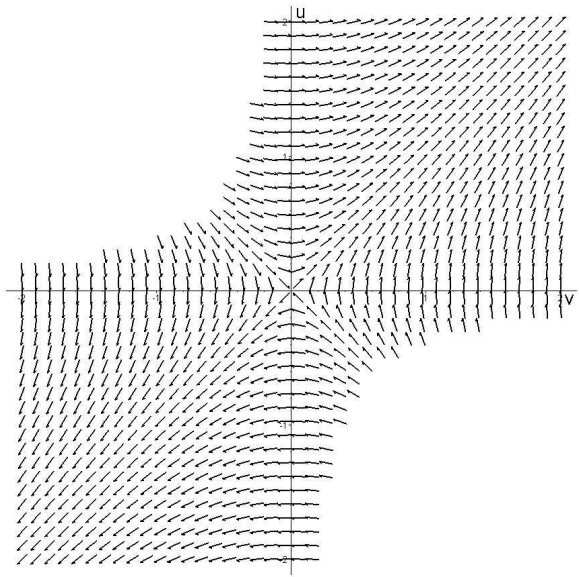,height=3in,width=3in,angle=0}
\caption{\label{schw2}The phase portrait for $k_{\alpha}$ for the Schwarzschild spacetime. The region below $r=0$ has been removed.}
\end{figure}

We now wish to classify the isolated critical point at the bifurcation 2 - sphere $u=v=0$. For the covariant Hessian (\ref{chessian}) we find
\begin{equation}\label{Huuvv}
    H_{uu}=\frac{\epsilon}{u^2}, \;\;\; H_{vv}=\frac{\epsilon}{v^2}
\end{equation}
where
\begin{equation}\label{epsilon}
    \epsilon \equiv \left( \frac{63}{2} \right)\frac{\mathcal{L}^2}{m^4(1+\mathcal{L})^{10}},
\end{equation}
\begin{equation}\label{Huv}
    H_{uv} = H_{vu} = \left( \frac{9}{2} \right) \left( \frac{\mathcal{L}}{uv} \right) \frac{(7\mathcal{L}-1)}{m^4(1+\mathcal{L})^{10}},
\end{equation}
and
\begin{equation}\label{Hthetatheta}
    H_{\theta \theta} = \frac{H_{\phi \phi}}{(\sin(\theta))^2} = -\left( \frac{9}{2} \right) \frac{\mathcal{L}}{m^4(1+\mathcal{L})^7}.
\end{equation}
The resultant determinant is given by
\begin{equation}\label{detHs4}
    H = -\left( \frac{9}{2} \right)^4 \left( \frac{\mathcal{L}^2}{uv} \right)^2 \frac{(14 \mathcal{L}-1)\sin(\theta)^2}{m^{16}(1+\mathcal{L})^{34}}.
\end{equation}
Noting that $\mathcal{L}(0)=0$ and that $\lim_{x \rightarrow 0}\mathcal{L}(x)^2/x = 0$ we conclude that
\begin{equation}\label{h00}
    H|_{\mathcal{P}} = 0
\end{equation}
and so the bifurcation two - sphere of the Schwarzschild solution is a degenerate critical point of the gradient flow when classified in the full spacetime. In a similar fashion one can show that the associated cofactors are also degenerate,
\begin{equation}\label{deltazero}
    \Delta_{n}|_{\mathcal{P}} =0, \;\;\; n=1,...,4.
\end{equation}
Since the flow is restricted to the $u-v$ plane, we construct the Hessian $\tilde{H}$ there. We find
\begin{equation}\label{hessian2}
    \tilde{H} = \left(\frac{9}{2m^4}\right)^2\left(\frac{\mathcal{L}}{uv}\right)^2\left(\frac{14\mathcal{L}-1}{(1+\mathcal{L})^{20}}\right)
\end{equation}
and so
\begin{equation}\label{hessian2}
    \tilde{H}|_{\mathcal{P}} = -\left(\frac{9}{2m^4}\right)^2.
\end{equation}
The critical point at the bifurcation 2 - sphere is a saddle point. The function $m^4\mathcal{I}$ is shown in Figure \ref{schw1}. The phase portrait for $k_{\alpha}$ is shown in Figure \ref{schw2}.
\subsection{Spherically Symmetric Spacetimes}
 Now \textit{locally}, every spherically symmetric spacetime can be written in the form
\begin{equation}\label{ssgen}
    ds^2=2f(u,v)dudv+r(u,v)^2 d\Omega^2_{2}.
\end{equation}
As above in Schwarzschild we consider the negative gradient flow of the square of the Weyl tensor. It follows that
\begin{equation}\label{w1ssgen}
    \mathcal{I} = \left(\frac{4}{3}\right)\frac{\kappa(u,v)}{f^6r^4},
\end{equation}
\begin{equation}\label{kssgen}
    k^{\alpha} = -\left(\frac{8}{3}\right)\frac{\kappa}{f^8r^5}(\zeta(u,v),\eta(u,v),0,0),
\end{equation}
and
\begin{equation}\label{KKssgen}
     \mathcal{V} = -\left(\frac{128}{9}\right)\frac{\kappa^2 \zeta \eta}{f^{15}r^{10}}.
\end{equation}
$\kappa$, $\zeta$ and $\eta$ involve derivatives of $f$ and $r$ up to order $p=2$. The explicit forms are not needed here. Rather, we note that there exist two distinct types of critical points associated with the gradient flows of $\mathcal{I}$. From (\ref{kssgen}) it follows that the critical points can be characterized by the \textit{local} conditions
\begin{equation}\label{critssgen}
    \kappa=0, \;\;\; \zeta = \eta = 0.
\end{equation}
In the Robertson-Walker case, of course, $\kappa=0$ is a global condition and so we examined Ricci invariants as explained above. The Schwarzschild spacetime demonstrates a case where $\kappa \neq 0$ near $\mathcal{P}$. The bifurcation 2-sphere corresponds to an isolated critical point characterized by $\zeta|_{\mathcal{P}} = \eta|_{\mathcal{P}} = 0$. We call such critical points locally \textit{anisotropic}. If $\kappa|_{\mathcal{P}} = 0$ we refer to such points as locally \textit{isotropic}. Locally isotropic critical points could be false critical points in the sense that their critical nature could derive from the use of quadratic (or higher) order invariants. This is examined below.

In order to \textit{explicitly} demonstrate examples where locally isotropic critical points are \textit{not} false critical points, consider, for example, static perfect fluids. In comoving coordinates we have
\begin{equation}\label{perff}
    ds^2=\frac{dr^2}{1-\frac{2 m(r)}{r}}+r^2 d\Omega^2_{2}-e^{\Phi(r)}dt^2.
\end{equation}
Given $\Phi$ sufficiently smooth, and subject to boundary conditions, one can generate the effective gravitational mass $m(r)$ (see the Appendix) that gives rise to a perfect fluid \cite{lake1}. For example, setting
\begin{equation}\label{perfs}
    \Phi=\frac{1}{2}N \ln(1+\frac{r^2}{\alpha}),
\end{equation}
where $N$ is an integer $\geq 1$ and $\alpha$ is a constant $>0$, an infinite number of exact perfect fluid solutions follow. Again, taking $\mathcal{I} = 8 w1R$, for all such models we have
\begin{equation}\label{center}
    k^{\alpha} = \mathcal{I}=\mathcal{V}=\square \;\mathcal{I}=0
\end{equation}
at the center of symmetry $r=0$. We have a critical timelike 3-surface at the origin based on the Weyl invariant. That is, the solutions are locally isotropic about the origin for all $t$. Defining $\mathcal{I} = (8 w1R)^{1/2}$ or, say, $\mathcal{I} = (8 w1R)^{1/4}$, we arrive back at (\ref{center}) at the origin.
\section{Discussion}
This paper serves as an introduction to the study of gradient flows of scalar invariants as a means to visualize curvature. No particular observer, or theory, is fundamental to this approach, but the intention is to apply the techniques to spacetimes associated with Einstein's theory of gravity. This allows for a physical understanding of the scalars in use. For example, a physical understanding of the Ricci invariants follows directly from Einstein's equations. Because the approach used here does not single out particular observes, we have shown that this restriction limits the number of invariants that can be used to construct gradient flows. Only the 4 Ricci invariants and the 4 Weyl invariants (in terms of its electric and magnetic components) can be used as building blocks to study gradient flows in the most general case. We have shown that higher order (mixed) invariants explicitly exhibit vectors and scalars associated with the timelike 4-vector used to split the Weyl tensor.

It has been shown that gradient flows of invariants polynomial in the Riemann tensor are necessarily orthogonal to Killing flows, should they exist. (This allows for a rigorous definition and generalization of the classical notions of $\mathcal{R}$ and $\mathcal{T}$ regions of spacetime.) From the point of view of dynamical systems, gradient flows are simple in the sense that critical points are the only possible limit sets of the flow. However, the classification of critical points must be made in a coordinate independent fashion and so a covariant classification scheme was developed. This, along with the Poincar\'{e} index, classify the ``topology" of the flow essential to the visualization.

For purely electric Ricci - flat spacetimes one can construct strict Newtonian analogues, based on the topology of flows; for the first electric (tidal) invariant in spacetime, and for the tidal invariant in Newtonian theory. This is discussed at length elsewhere \cite{curzon}.

We have reviewed examples, restricted here to spherical symmetry. A rather complete analysis of the Kruskal - Szekeres vacuum was given, interpreting the associated bifurcate 2-sphere as the isolated critical point of the solution. More generally, isolated critical points of the Weyl invariant within spherical symmetry were distinguished as locally isotropic or anisotropic, with explicit examples given for each.

\bigskip

\begin{acknowledgments}
This work was supported in part by a grant from the Natural Sciences and Engineering Research Council of Canada. Portions of this work were
made possible by use of \textit{GRTensorII} \cite{grt}.
\end{acknowledgments}

\appendix*
\section{$\mathcal{R}$ and $\mathcal{T}$ regions}

The gradient of the areal radius of an arbitrary spherically
symmetric field can be spacelike, timelike or null. A careful
distinction of these possibilities is important and, for example,
forms an essential element of any complete proof of the Birkhoff
theorem (for example \cite{he}). The possibility of spacelike and
timelike gradients was studied extensively in the Russian
literature (and labeled ``$\mathcal{R}$" and ``$\mathcal{T}$" regions respectively) some
forty years ago \cite{rt} and yet there appears to be no readily
available extension of these ideas. It is the purpose of this Appendix to explore this using gradient flows.

The original distinction of $\mathcal{R}$ and $\mathcal{T}$ regions of spacetime was
restricted to spherical symmetry and given by Novikov \cite{rt}
(see also the work of Ruban \cite{rt}) by way of a coordinate
construction \cite{zeld}. This construction can be made invariant
as follows: Consider a spherically symmetric spacetime
\begin{equation}
ds^2=ds^2_{\Sigma}+R^2d\Omega_{2}^2
\end{equation}
where $d\Omega_{2}^2$ is the metric of a unit 2-sphere
($d\theta^2+\sin^2(\theta)d\phi^2$) and $R=R(x^1,x^2)$ (the areal
radius) where the coordinates on the Lorentzian 2-space $\Sigma$
are labeled as $x^1$ and $x^2$. (No specific choice of
coordinates on $\Sigma$ need be made.) Writing the effective
gravitational mass (the Misner-Sharp energy) as
$m=m(x^1,x^2)$\footnote{The function $m$ is invariantly defined, the
simplest explicit formula being $m=
\frac{R^{3}}{2}R_{\theta \phi}^{\; \; \;\; \theta \phi}$. Further background
information can be found in \cite{mass}} it follows from Novikov's coordinate
definitions that
\begin{equation}
R> 2m, \;\;\;\;R=2m,\;\;\;\;R<2m \label{apparent}
\end{equation}
in an $\mathcal{R}$ region, on the boundary and in a $\mathcal{T}$ region
respectively \cite{mcvittie}. In more modern notation
\cite{hayward}, an $\mathcal{R}$ region is untrapped, a $\mathcal{T}$ region trapped
and the boundary marginal.  Whereas these distinctions are
fundamental, they are restricted to strict spherical symmetry.

In order to bring gradient flows into the picture, let us note that a stationary region of spacetime admits a timelike Killing congruence. Since every non-zero 4-vector orthogonal to a timelike 4-vector must be spacelike, it follows from (\ref{killing}) that any gradient flow is necessarily spacelike in a stationary region. From (\ref{norm}) we have $\mathcal{V}>0$ throughout an $\mathcal{R}$ region, a fact that we can use to define such a region. Now let us assume that there is a region in which the gradient flow is timelike so that $\mathcal{V}<0$. (A Killing congruence in this region, should one exist, is then necessarily spacelike.) We can use $\mathcal{V}<0$ to define a $\mathcal{T}$ region. Boundary regions are then naturally defined by $\mathcal{V}=0$. The foregoing classification is independent of symmetries but depends on one's choice of invariant used to construct the gradient flow. We adopt this classification in general, always referenced to a particular gradient flow: $\mathcal{V}>0$ throughout an $\mathcal{R}$ region where $k$ is spacelike, $\mathcal{V}<0$ throughout an $\mathcal{T}$ region where $k$ is timelike, and $\mathcal{V}=0$ on a boundary where $k$ is null or the zero vector. It is already known \cite{lake}, for example, that for the Kerr metric, $\mathcal{V} \geq 0$ for the rotational parameter $A \geq 1$ with equality holding only in the degenerate case $A=1$. That is, degenerate and naked Kerr metrics do not admit $\mathcal{T}$ regions.\footnote{ The Kerr metric has two independent (Weyl) invariants derivable from the Riemann tensor without differentiation. Both invariants give this same general result. However, the coordinate locations of the $\mathcal{R}$ and $\mathcal{T}$ regions differ depending on which invariant is chosen. See \cite{lake} for details.}


\begin{thebibliography}{}
\bibitem{kruskal}Note that our simple labeling of $u$ and $v$ differs from the usual labeling which uses $\mathcal{L}(-uv/e)$.
\end{thebibliography}
\end{document}